**CES Working Papers**

# Initial Crypto-asset Offerings (ICOs), tokenization and corporate governance

Stéphane BLEMUS, Dominique GUEGAN

**2019.04**

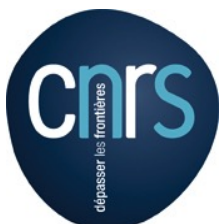

# Initial Crypto-asset Offerings (ICOs), tokenization and corporate governance

Stéphane Blemus* & Dominique Guégan**

* Paris 1 Panthéon-Sorbonne University, LabEx ReFi, Kalexius law firm, ChainTech

** Centre d'Economie de la Sorbonne, Paris 1 Panthéon-Sorbonne University, Labex ReFi, Ca'Foscari University, Venezia, IPAG Business School

This interdisciplinary paper by a mathematician and a legal counsel, both from the Paris 1 Panthéon-Sorbonne University, discusses the potential impacts of the so-called "initial coin offerings", and of several developments based on distributed ledger technology ("**DLT**"), on corporate governance. While many academic papers focus mainly on the legal qualification of DLT and crypto-assets, and most notably in relation to the potential definition of the latter as securities/financial instruments, the authors analyze some of the use cases based on DLT technology and their potential for significant changes of the corporate governance analyses.

This article studies the consequences due to the emergence of new kinds of firm stakeholders, i.e. the crypto-assets holders, on the governance of small and medium-sized enterprises ("**SMEs**") as well as of publicly traded companies. Since early 2016, a new way of raising funds has rapidly emerged as a major issue for FinTech founders and financial regulators. Frequently referred to as initial coin offerings, Initial Token Offerings ("**ITO**"), Token Generation Events ("**TGE**") or simply "token sales", we use in our paper the terminology Initial Crypto-asset Offerings ("**ICO**"), as it describes more effectively than "initial coin offerings" the vast diversity of assets that could be created and which goes far beyond the payment instrument issue.

An ICO can be summarized as follows: a new method to raise funds through the offer and sale by a group of developers or a company to a crowd (i.e. investors or contributors) of ad hoc crypto-assets (also coined as "tokens") specifically created and issued on a distributed ledger, sometimes preceded by an early sale of the crypto-assets called "pre-sale", for the purpose of launching a business or of developing ad hoc governance of projects based, in several cases, on the distributed ledger technology, typically in exchange for pre-existing 'mainstream' crypto-assets, such as Bitcoin and Ether among others, or even fiat currencies. Perceived by several entrepreneurs as a less burdensome way of fundraising, at least 25 billion dollars have been raised between March 2016 and August 2018 through ICOs only[1].

Beyond the strict debate on the ICO process, we are witnessing a paradigm shift at a broader spectrum through the development of these new kinds of assets, the crypto-assets, whose creation can be summarized in one word: tokenization. Indeed, crypto-assets could be created through different processes, and some of them have and will be created without the need to proceed to an ICO, e.g. Bitcoin which has been created through mining activity by a decentralized group of developers without any central third party involved and without any fundraising ambition, and Telegram which has raised enough funds during its pre-sale without the need and the will to organize its initially planned ICO.

---

[1] According to ICO statistics available on the website Coinschedule.com.

Each crypto-asset, issued or not during an ICO, will have its own underlying characteristics and will give rights to third parties which are specific for each project. Use cases of crypto-assets are extremely diverse, as some are created to grant rights akin to a currency, others resembling to an equity or to a debt instrument, and others similar to a utility asset which would allow its holder to have access to certain non-financial advantages. In addition, once they are issued, the crypto-assets may in principle be resold in a secondary market[2]. The legal status of "crypto-assets" (or "tokens"), whether they are utility tokens, security tokens and/or payment tokens, is still unclear and will be subject to several regulatory decisions in the months and years to come in the European Union ("**EU**"), in the United States of America ("**US**") and elsewhere.

After several studies on the legal issues related to the crypto-asset issuance, one of the main key problematic issues that will have to be handled by legal, economic and financial experts in the coming years is to understand the various outcomes of ICOs, of tokenization and, more globally, of distributed ledger technology on the governance of companies.

Corporate governance can be defined as the mechanisms which have an impact inside a company/firm on the relationship between the company's management, its board, its shareholders and other stakeholders, and whose aim is to provide an efficient structure to control and monitor effectively the company's management, objectives and performance. According to the principles defined by the G20/OECD, the main goal of corporate governance is "to support economic efficiency, sustainable growth and financial stability"[3].

ICOs, crypto-assets and tokenization, as well as more broadly the use of distributed ledger technology and of smart contracts, could impact the way corporates are governed in several ways. New technologies, and blockchain and crypto-assets in particular, have the potential to transform basic notions fundamental to corporate governance, such as trust, intermediation, accountability, responsibility and transparency[4]. Indeed, distributed ledger technology can, at least in theory, reduce the cost of accessing information for minority shareholders/stakeholders and enhance transparency in the governance of firms. This new technology could modify risk management, compensation governance, accountability to shareholders, and redefine the current shift between stakeholders and shareholders. The positive effects of the distributed ledger technology could be substantial, and so are the various new legal and economical challenges it represents for companies, among which: are crypto-asset holders a new kind of corporate stakeholders? Could they be considered as shareholders? Or as bondholders? How could they participate in the governance of the company? Are smart contracts new tools for corporate governance? How to solve free-rider problems in the process of ICOs?

---

[2] With the exception of the non-convertible/closed crypto-assets which are designed and "intended to be specific to a particular virtual domain or world (e.g. Q Coins), and under the rules governing its use". Source: The Financial Action Task Force ("**FATF**"), "Virtual Currencies – Key Definitions and Potential AML/CFT Risks", June 2014.
[3] OECD, "G20/OECD Principles of Corporate Governance", OECD Publishing, 2015, page 9.
[4] M. Fenwick and E. Vermeulen, "Technology and Corporate Governance: Blockchain, Crypto and Artificial Intelligence", European Corporate Governance Institute, ECGI Working Paper No. 424/2018, November 2018; V. Akgiray, "Blockchain Technology and Corporate Governance", Report for the OECD Corporate Governance Committee's roundtable discussion on blockchain technologies and possible implications for effective use and implementation of the G20/OECD Principles of Corporate Governance, 6 June 2018.

More broadly, and as computer code will progressively be used in conjunction with – and potentially sometimes replace – written legal codification, the key issue is to analyze if it would be feasible and advisable to govern an organization, e.g. a company, in a decentralized and distributed way. While the ICO of the "The DAO" project partially failed in 2016 in a dramatic way, some scholars are looking beyond it and envisioning new types of corporations and institutions, i.e. distributed autonomous organizations, which would be managed at least partially through autonomous codes and which would provide a more horizontal, distributed and/or algorithmic governance based on the use of self-executing smart contracts.

In this interdisciplinary paper, the authors examine in depth the opportunities and risks raised by ICOs, tokenization and distributed ledger technology in terms of corporate governance. First the concepts of ICO and tokenization, and beyond the consequences of distributed ledger technology, will be discussed in the context of a corporate governance perspective; secondly the potential rights granted by different types of crypto-asset holders (whether utility tokens or security tokens) will be carefully analyzed; thirdly the possible governance of a decentralized and distributed organization trying to answer the previous questions will be addressed.

# Table of contents

**I. Introduction to Initial Crypto-assets Offerings (ICOs) and the tokenization process**

a. <u>Defining ICOs</u>

In order to apprehend the impact of ICOs on corporate governance, defining at first the key terms and concepts used in the paper is essential, such as: (i) distributed ledger technology, (ii) smart contracts and (iii) initial token/crypto-asset offerings.

    i. What is distributed ledger technology (DLT)?

A distributed ledger (also called shared ledger, or distributed ledger technology, or DLT) is a mathematically secured, chronological and decentralized ledger[5] of digital data replicated, shared/distributed and maintained accessible to a network of computers/nodes connected on a peer-to-peer basis, based on a consensus mechanism of technological validation, such that network participants can share and retain identical, cryptographically secured and immutable data records in a decentralized manner[6].

One of the key characteristics is that the data distributed through a distributed ledger is maintained by its participants, and not by a central database administrator or party. There is no centralized data storage. Every network participant can have an identical copy of the relevant distributed ledger. Based on a consensus mechanism and encrypted technology, additions to the database such as new transactions are grouped together and validated by a network of participants ("nodes"). A peer-to-peer network is required as well as consensus algorithms to ensure replication across nodes.

There are three main different operating models for DLTs: public DLTs; consortium DLTs; and fully private DLTs[7]. Public DLTs are distributed ledgers where anyone can participate through the consensus process and interact without any access restrictions. The opposite model is the fully private DLT where participants are part of a single organization and which could be applied for auditing and internal management purposes. And between these two models, consortium or "partially decentralized" DLTs are distributed ledgers where the "consensus process is controlled by a pre-selected set of nodes."[8]

Probably the most prominent distributed ledger technology nowadays is known as the "**Blockchain**", which has been used as the underlying technology of the Bitcoin crypto-currency. Transactions registered on a Blockchain are aggregated in "blocks" and appended to existing records in a decentralized network or "chain" (hence the name blockchain). An encrypted signature is used to validate any transaction.

---

[5] The first part of this definition of "blockchain" has been adopted in the US State of Vermont Statutes of 2016 (Rule of Evidence, Title 12, Chapter 81, §1913) and in the US State of California Assembly Bill No. 2658 chaptered on 28 September 2018.
[6] S. Blemus, "Law and Blockchain: A Legal Perspective on Regulatory Trends Worldwide", Revue Trimestrielle de Droit Financier (Corporate Finance and Capital Markets Law Review), RTDF n°4-2017, December 2017.
[7] V. Buterin, "On Public and Private Blockchains", Ethereum Blog, 7 August 2015; D. Guegan, "Public Blockchain versus Private Blockchain", Documents de travail du Centre d'Economie de la Sorbonne, 2017.
[8] Ibid.

While some people consider that a Blockchain is synonymous with a distributed ledger (and generalist media tends to use it with this meaning), others argue that technically it would only apply to linear distributed ledgers such as the ones Bitcoin or Litecoin use and not to Directed Acyclic Graphs (DAG) such as the ledgers based on Iota, Tangle or Hedera Hashgraph algorithms. Therefore, according to the latter definition, not all distributed ledgers have to necessarily employ a chain of blocks to successfully provide secure and valid achievement of distributed consensus: a Blockchain is only one type of data structure considered to be a distributed ledger.

Thus, in the following article, we will keep the more general definition of "DLT", considering that in a distributed ledger system, each member of the network (or node) holds a localized, synchronized copy of the ledger, and that DLT is characterized by resilience and immutability, as tampering or destroying one node does not affect the entire system, nor the continuity of the ledger. DLT can bring a number of benefits, including building authentication and tamper proof history.

ii. What is a smart contract?

With the development of the distributed ledger technology, parties can therefore now have a shared database, a common mapping of events and transactions in real time. They can also provide the self-execution of certain pre-determined outcomes when certain events arise, the so-called "**smart contracts**".

Indeed, among the many promising features now possible with the existence of DLT, the self-executing and automatable computer programs known as "smart contracts" appear to be one of the most disruptive and challenging developments. Albeit frequently used as a component of DLT, smart contract is a specific technology and a subject of analysis per se. The first detailed description of "smart contract" was written by cryptography pioneer Nick Szabo in 1994: "a computerized transaction protocol that executes terms of a contract. The general objectives of smart contract design are to satisfy common contractual conditions (such as payment terms, liens, confidentiality, and even enforcement), minimize exceptions both malicious and accidental, and minimize the need for trusted intermediaries. Related economic goals include lowering fraud loss, arbitration and enforcement costs, and other transaction costs."[9]

There is no unanimous definition for "smart contracts", mainly because computer and legal experts have different perceptions of what a contract is. Law professionals define a contract as a formal legally binding agreement between parties, while computer engineers perceive it as computer code, i.e. an arrangement of data and computer instructions executing pre-selected actions in computer programs.

To summarize what is a "smart contract", an ISDA[10] report published in August 2017[11] quotes the Clack, Bakshi and Braine definition: "a smart contract is an automatable and enforceable agreement. Automatable by computer, although some parts may require human input and control. Enforceable either by legal enforcement of rights and obligations or via tamper-proof execution of computer code."[12] The fundamental elements of "smart contracts" are the following ones in the definition we propose: an agreement existing in the form of computer code which provides the automation and the self-execution of a pre-determined action upon fulfillment of pre-determined conditions.[13]

---

[9] N. Szabo, "Smart Contracts", 1994. See also: N. Szabo, "Formalizing and Securing Relationships on Public Networks", First Monday, 1st September 1997; K. Werbach and N. Cornell, "Contracts ex machina", Duke Law Journal, 2017.
[10] International Swaps and Derivatives Association Inc., or "ISDA".
[11] ISDA and Linklaters, "Whitepaper on Smart Contracts and Distributed Ledger - A Legal Perspective", pages 4-5, August 2017.

As for its use cases, the UK government chief scientific adviser advised in early 2016 that its "potential benefits include low contracting, enforcement, and compliance costs, while potential risks include a reliance on the computing system that executes the contract".[14] Smart contracts will notably be used to allow business processes and data validation in a shared, tamper-proof and resilient space, especially related to the transfer of crypto-assets on distributed ledgers. In that respect, in a bill signed into law by the US State of Arizona in March 2017, smart contracts are defined as such: "an event-driven program that runs on a distributed, decentralized, shared, and replicated ledger that can take custody over, and instruct transfer of, assets on that ledger".[15]

   iii. What is an ICO and what are crypto-assets?

Beyond the creation of the Bitcoin cryptocurrency, the rise of distributed registry technology has led to the emergence of ecosystems that allow the provision of new services even though they are still often experimental. These are complex technological projects aimed at a tech-savvy public likely to understand this new type of environment, even though the general public will progressively be aware of these developments.

Among these new ecosystems, since 2016 a new means of fundraising has been developed which uses the DLT technology. Initial Crypto-asset Offerings (ICOs) (also called "Token Generation Event" or "tokens sales") can be described as new ways to raise funds through the use of DLT, resulting in the issuance of cryptographic tokens (or "crypto-assets" or "tokens").

Crypto-assets are intangible, math-based, digital and cryptographically-secured assets, issued, registered, retained or transferred through cryptography and, in most cases, DLT, representing a crypto-asset holder's rights to receive a benefit or perform specified functions. Crypto-assets issued during an ICO are initially transferred to investors through smart contracts' computer code, in exchange of a pre-existing cryptocurrency or against a fiat currency such as US dollars and euros, for instance, at a price set by the token issuer[16] to finance a project, then potentially

---

[12] C. Clack, V. Bakshi and L. Braine, "Smart Contract Templates: foundations, design landscape and research directions", Cornell University Library, 2016 and revised in March 2017.
[13] About the definition of smart contracts, see also: S. Blemus, "Law and Blockchain: A Legal Perspective on Regulatory Trends Worldwide", Revue Trimestrielle de Droit Financier (Corporate Finance and Capital Markets Law Review), RTDF n°4-2017, December 2017; A. Savelyev, "Contract Law 2.0: "Smart" Contracts As the Beginning of the End of Classic Contract Law", Higher School of Economics Research Paper, SSRN, 14 December 2016; M. Raskin, "The Law and Legality of Smart Contracts", Georgetown Law Technology Review, Volume 1, 25 September 2016, p. 306; H. Surden, "Computable contracts", UC Davis Law Review, 2012, pp. 629-700.
[14] M. Walport, Government Office for Science, "Distributed Ledger Technology: beyond block chain", Report by the UK Government Chief Scientific Adviser, 19 January 2016, cited in: A. Delivorias, "Distributed ledger technology and financial markets", European Parliamentary Research Service, Briefing, November 2016.
[15] Arizona House Bill n°2417, signed into law on 29 March 2017. The smart contract definition has been added in Title 44, chapter 26, article 5(E)(1) of the Arizona Revised Statutes. The US State of Tennessee passed a similar law, the Public Chapter No. 591 of the Tennessee Public Acts of 2018, on 26 March 2018.
[16] Token issuers can be either private or public entities. See: Deloitte, "State-Sponsored Cryptocurrency: Adapting the best of Bitcoin's Innovation to the Payments Ecosystem", 2015; J. Barrdear and M. Kumhof, "The

tradable on a secondary market (on platforms such as Kraken and Poloniex). It is to be noted that crypto-assets can also be created without an ICO by issuing them through other kinds of rewards (e.g. the mining model[17]), and not by the issuance by a single organization, such as for Bitcoin.

The emergence of ICOs enables entrepreneurs to respond to two fundamental needs of the DLT ecosystem: (i) the creation of incentive mechanisms to participate to this ecosystem and/or to innovate (through the granting of financials rights, utility rights…), and (ii) the financial ability to fund projects related to the new innovative distributed ledger technology (the first entrepreneurs for DLT projects did not find adequate sources of financing). The ICO developers are supposed to raise this capital to fund their digital platform, software or other projects at an early stage of their development.

In comparison to traditional asset categories such as financial securities (e.g. debt or equity), crypto-assets can have a plurality of functions. Besides, each type of crypto-asset issued during an ICO possesses its own underlying characteristics, granting rights different from the other crypto-assets (voting rights, share of capital or any particular advantage ...). This makes defining what a token is difficult from a regulatory perspective. Use cases of crypto-assets have evolved way beyond just being a virtual (or crypto) currency. A token can be used with the attributes of an equity instrument (e.g. conferring immediate or future ownership or equity interest in the legal entity, voting rights and/or a promise to share in future profits and/or losses), of a debt instrument (e.g. granting the right to hold the status of a creditor/lender), or can be designed as a utility services asset giving the right to access or license a service/product or to use, sell or consume the item purchased. This diversity of use cases is resumed by Richard Olsen as such: "there won't be millions of tokens. There will be millions of kinds of tokens." Besides, once they are issued, the crypto-assets may be resold in a secondary market, through brokers, exchange platforms or over-the-counter transactions.

Taxonomies of tokens have already been proposed by a substantial number of studies published by crypto-entrepreneurs, academics and regulators worldwide[18]. In this paper, our proposal is to classify the existing tokens under three main categories: (i) the security tokens; (ii) the utility tokens; and (iii) the crypto-currency tokens:

- Security tokens (frequently referred to as "tokenized financial instrument"): this category includes tradable tokens whose primary purpose is to give holders voting or financial rights. It notably captures cases where tokens represent rights similar to the ones of a security, i.e. a debt or an equity instrument. Its definition is currently different among jurisdictions. For example, the US federal regulators have a broader definition

---

macroeconomics of central bank issued digital currencies", Bank of England, Staff Working Paper No. 605, July 2016; M. Bech and R. Garratt, "Central bank cryptocurrencies", BIS Quarterly Review, 17 September 2017.

[17] D. Guégan and C. Hénot, "A Probative Value for Authentication Use Case Blockchain", Documents de travail du Centre d'Economie de la Sorbonne, 2018 ; D. Guégan and A. Sotiropoulou, "Bitcoin and the challenges for financial regulation", Capital Markets Law Journal, Volume 12, Issue 4, October 2017, pp. 466-479.

[18] Financial Conduct Authority, English HM Treasury and Bank of England, "Cryptoassets Taskforce: final report", October 2018, pp. 11-14; K. Bheemaiah and A. Collomb, "Cryptoasset valuation: Identifying the variables of analysis", Louis Bachelier Institute, Working Report v1.0, 19 October 2018; K. Lachgar and J. Sutour, CMS Bureau Francis Lefebvre, "Le token, un objet digital non identifié?", Option Finance, 13 November 2017 ; G.J. Nowak and J.C. Guagliardo, Pepper Hamilton LLP, "Blockchain and initial coin offerings : SEC provides first US securities law guidance", Harvard Law School Forum on Corporate Governance and Financial Regulation, 9 August 2017 ; H. de Vauplane, Kramer Levin Naftalis & Frankel LLP, "Quelle régulation pour les offres publiques en cryptomonnaies (ICO)?", Revue Banque, 28 June 2017.

of what constitutes a security (based on an extensive interpretation of the so-called "Howey" test)[19] than the French or than the Swiss financial regulators[20]. According to Tapscott, there is six types of security tokens: "debenture tokens; smart swap contracts; option tokens; equity tokens; bond tokens; and smart futures contracts."[21]

- Utility tokens[22]: this category includes tokens construed as conferring rights to access to an application or a service or a product, or to use, sell or consume the item purchased, and that generally require the use of a blockchain-type infrastructure. Examples: Golem (GNT), Augur (REP) and SiaCoin (SC).

- Crypto-currency tokens (synonymous with "payment tokens" or "money instrument tokens"): this category includes tokens accepted as a means of payment for the purchase of goods and/or services, or to be used for the money or value transfer[23]. Examples: Bitcoin (BTC), Bitcoin Cash (BCH), Litecoin (LTC) and Monero (XMR).

b. <u>Defining ICOs rationale and impact</u>

i. Economic rationale for the creation of ICOs and crypto-assets

The reasons behind the creation of crypto-assets have evolved throughout the years. The first crypto-asset created, the Bitcoin, was not made through an ICO, but via mining activity. The foundation of the Bitcoin *raison d'être*, according to the famous paper published in 2008 in the aftermath of the systemic financial crisis by Satoshi Nakamoto, widely regarded as the creator of the Bitcoin protocol, consists in "an electronic payment system based on cryptographic proof instead of trust, allowing any two willing parties to transact directly with each other without the need for a trusted third party."[24] The monetary issue is key in Nakamoto's reasoning, and not far from what the economist Milton Friedman predicted as soon as in 1999: "the one thing that's missing [on the Internet], but that will soon be developed, is a reliable e-cash, a method whereby on the Internet you can transfer funds from A to B without A knowing B or B knowing A."[25]

---

[19] The current position of the US federal regulator Securities and Exchange Commission (SEC) is based on the interpretation of the term "security" as defined in US Supreme Court decisions such as US Supreme Court, "Securities and Exchange Commission v. W.J. Howey Co.", 328 US 293, 298-99 (27 May 1946); US Supreme Court, "United Housing Foundation Inc. v Forman", 421 US 837 (1975); and US Supreme Court, "Securities and Exchange Commission v. Charles E. Edwards", 540 US 389, No. 02-1196 (27 May 2004). See: "Report of Investigation under 21(a) of the Securities Exchange Act of 1934: The DAO", Release No. 81207, and "Investor Bulletin: Initial Coin Offerings", 25 July 2017; R. B. Levin and P. Waltz, "The devil is in the details: SEC regulation of Blockchain technology", Polsinelli eAlert, March 2017.

[20] Several terminologies are used by national regulators to define security tokens: The Swiss financial regulator FINMA uses the term "asset tokens", the US SEC regulator has used recently the wording "digital asset securities", and the French financial regulator ("Autorité des Marchés Financiers" or "**AMF**") prefers the following wording: "tokens offering political or financial rights". Cf. SEC, "Statement on Digital Asset Securities Issuance and Trading", 16 November 2018; Directorate General for Economic Development, Research and Innovation (DG DERI) of the State of Geneva, "Guide: Initial Coin Offerings (ICOs) in the Canton of Geneva", 28 May 2018; AMF, "Summary of replies to the public consultation on Initial Coin Offerings (ICOs) and update on the UNICORN Programme", 22 February 2018; FINMA, "Guidelines for enquiries regarding the regulatory framework for initial coin offerings (ICOs)", 16 February 2018.

[21] A. Tapscott, "Taxonomy of crypto-assets", speech during "Consensus 2018", 4th annual blockchain technology summit, May 2018.

[22] According to article 26 of the French "PACTE" draft bill, a crypto-asset (or token) is characterized as "any intangible asset which represents, in digital form, one or more rights which can be issued, registered, retained or transferred through a [distributed ledger technology] that would allow to identify, directly or indirectly, the owner of the asset." Source: Government-sponsored draft bill n°179 adopted by the French National Assembly on 9th October 2018 related to companies' growth and transformation.

[23] J-P. Landau, "Les crypto-monnaies", report to the French Minister for Economy and Finance, 4 July 2018.

[24] S. Nakamoto, "Bitcoin: A Peer-to-Peer Electronic Cash System", White paper, 2008. See also: G. Selgin, "Synthetic Commodity Money", SSRN, 6 February 2012.

After the creation of Bitcoin, the various crypto-assets created since then have been issued to solve many issues way beyond the only currency problematic, and the creation of a new kind of fundraising called ICO has helped in that matter.

In 2016, the amount raised in ICOs was 98,6 million dollars. In 2017, it climbed to 6,2 billion dollars, and in the eight first months of 2018 alone the figure reached 18,7 billion dollars[26]. In less than two years of existence, ICOs and crypto-assets issuances have eclipsed venture capital as the primary source of funding for early-stage tech companies.

According to Ennis, Waugh & Weaver, the fundamental reasons behind this tremendous growth related to ICO fundraisings could be summarized in three fundamental reasons: "(1) a self-funding within the crypto-economy, (2) the deployment of a token within the ecosystem of an ICO project, and (3) the set of all economic activity generated through the creation of tokens."[27]

The main explanation behind ICO emergence was at first the difficulty of the early pioneering blockchain entrepreneurs to find investors. As these entrepreneurs could not raise enough capital through traditional fundraising models, innovative ways of fundraising were necessary. In the continuity of the disintermediation paradigm, ICOs were then invented to create a more direct relationship between DLT funders and investors, to attribute out-of-the-box and tailor-made rights to investors especially designed to the proposed project, and to permit DLT start-ups to develop without the existing intermediaries.

As time goes by, the scope of use cases of ICOs have further been expanded. The ICO developers are raising this capital, in most cases, to fund their digital platform, software or other projects based on the DLT technology at an early stage of their development. But, more globally, what we are witnessing is the tokenization of the economy and of many kinds of existing assets.

ii. The emergence of tokenization and tokenomics

The importance of the ICO and crypto-issuance development is to be understood through the economic functions of crypto-assets, what is frequently coined as "tokenomics". According to Mougayar, from an economical perspective, a crypto-asset is "a unit of value that an organization creates to self-govern its business model, and empower its users to interact with its products, while facilitating the distribution and sharing of rewards and benefits to all of its stakeholders."[28]

As explained previously, the first step to token creation has been to use the distributed ledger technology to address financing gaps (as applying regulation and fundraising models were only emerging). While the amount raised through ICOs has progressively surpassed early stage venture capital funding[29], and while creativity develops at a fast pace, the second step of token creation has been the development of several use cases based on the digitalization of all kinds of assets. Tokens can be created to develop a link between the token and an economic/non-economic ecosystem, and also can be created to record and transfer existing assets (such as commodities, securities…), and even illiquid asset classes (private equity & venture capital funds, antiques, art, oil and gas drilling, real estate, infrastructure, private unlisted debt, hedge funds, structured credit…), through computer code on the DLT. Beyond ICOs, the token creations which could be achieved through ICOs (and/or by pre-sales) or by the transformation of existing assets in the digital world accelerate a trend which is the digitalization of value and of goods. In theory, and without taking account of regulatory issues, each type of asset could be digitally represented by a token. This Internet of Things (IoT) perspective could further boost the dynamics of a new kind of economics called "tokenomics".

---

[25] P-E. Gobry, "Milton Friedman Predicted the Rise of Bitcoin in 1999", Forbes, 20 January 2014.
[26] "Cryptocurrency ICO Stats" for 2016, 2017 and 2018 published on Coinschedule.com.
[27] P.J. Ennis, J. Waugh & W. Weaver, "Three Definitions of Tokenomics", CoinDesk, March 2018.
[28] W. Mougayar, "Tokenomics – A Business Guide to Token Usage, Utility and Value", Medium, June 2017.
[29] CNBC, "Initial coin offerings have raised $1.2 billion and now surpass early stage VC funding", August 2017.

Ennis, Waugh and Weaver have rightly proposed three definitions of tokenomics: "(1) a means of self-funding within the crypto economy, (2) the deployment of a token within the ecosystem of an ICO project and (3) the set of all economic activity generated through the creation of tokens"[30]. In its primary meaning, the authors understand "tokenomics" as "a self-funding mechanism for projects within the crypto economy." A second definition of "tokenomics" would apprehend the notion of token, beyond this strictest definition, along the lines of its function, given that the token is an incentive to use the technology provided by the token issuer. Whether it is a utility token, a security token or a payment token, a token has an impact on the ecosystem of an ICO project, through its utility within the internal ecosystem of the issuance project as well as through its impact on its organization, e.g. the payment of salaries, goods and services useful for the project development. The authors describe also a third broader definition of tokenomics which focuses on the economic activity and value generated through the token creation.

ICOs and tokenization are topics only two years old, and therefore it is impossible to determine the full spectrum of all the use cases based on crypto-assets. We are only witnessing now to the potential digitalization of all kinds of existing "financial or tangible"[31] assets and of the creation of new kinds of rights which raise substantial governance issues. Security tokenization could include change in "the way in which regulation applies"[32], "fractionalization of larger assets, increased liquidity, lower issuance fees, and greater market efficiency", as well as providing to issuers an "access to a global pool of capital"[33] with "lower fees, more market exposure as deals are public and so visible to everyone with internet connection"[34].

## II. The disruption of corporate governance by ICOs and DLTs

Part II of this paper studies two potential important consequences of DLTs and ICOs on corporate governance: (i) how distributed ledger technology is the continuation of a longer process to digitalize the way corporates are governed, and (ii) how the potential role of the new corporate stakeholders, i.e. the token holders, could affect the balance of power within firms traditionally studied in corporate governance papers.

a. <u>Distributed ledger technology as the continuity of the digitalization of corporate governance</u>

---

[30] P.J. Ennis, J. Waugh and W. Weaver, "Three Definitions of Tokenomics", Coindesk, 17 March 2018 (https://www.coindesk.com/three-definitions-tokenomics).
[31] Financial Conduct Authority, English HM Treasury and Bank of England, "Cryptoassets Taskforce: final report", October 2018, p. 13.
[32] Ibid, p.13: "for example, there may be differences in the systems and controls that a firm needs to have."
[33] T. Koffman, "Your official guide to the security token ecosystem", Medium, 13 April 2018.
[34] H. Marks, "The future of US securities will be tokenized", Medium, 22 May 2018.

i. Understanding corporate governance basics

There are several definitions of corporate governance, and not one generally accepted definition. Corporate governance is a "multifaceted subject"[35] and an "ambiguous concept."[36] The traditional concept of corporate governance focuses on the structure of the company, referring "to the way in which a corporation is directed, administered, and controlled." "Corporate governance deals with the decision-making at the level of the board of directors and top management (i.e., the management board in a two-tier system), and the different internal and external mechanisms that ensure that all decisions taken by the directors and top management are in line with the objective(s) of a company and its shareholders, respectively."[37]

But "corporate governance also concerns the relationships among the various internal and external stakeholders involved as well as the governance processes designed to help a corporation achieve its goals."[38]

The OECD has published since 1999 "Principles of Corporate Governance"[39], which have become a benchmark on the issue, highlights a broader view: "Corporate governance involves a set of relationships between a company's management, its board, its shareholders and other stakeholders. Corporate governance also provides the structure through which the objectives of the company are set, and the means of attaining those objectives and monitoring performance are determined. Good corporate governance should provide proper incentives for the board and management to pursue objectives that are in the interests of the company and its shareholders and should facilitate effective monitoring."

The G20/OECD corporate governance principles of 2015[40] provide guidance on six main chapters: (i) ensuring the basis for an effective corporate governance framework; (ii) protecting the rights (to information and participation) and equitable treatment of shareholders and key ownership functions; (iii) promoting sound economic incentives throughout the investment chain and its numerous intermediaries (institutional investors, stock markets and other intermediaries); (iv) recognizing the role and rights of stakeholders and encouraging active co-operation between corporations and stakeholders; (v) ensuring a timely and accurate disclosure and transparency regarding the corporation; and (vi) ensuring the responsibilities of the board (the strategic guidance, the effective monitoring of management by the board and the board's accountability to the company and the shareholders).

As explicated by its current Secretary General Angel Gurria, "good corporate governance is not an end in itself. It is a means to create market confidence and business integrity, which in turn is essential for companies that need access to equity capital for long term investment. Access to equity capital is particularly important for future oriented growth companies and to balance any increase in leveraging."[41]

---

[35] H. Kent Baker and R. Anderson, "An Overview of Corporate Governance", in "Corporate governance: a synthesis of theory, research and practice", 2010, p. 15.
[36] A. Keay, "The Enlightened Shareholder Value Principle and Corporate Governance", 2013.
[37] P. O. Mülbert, "Corporate Governance of Banks after the Financial Crisis – Theory, Evidence, Reforms", European Corporate Governance Institute (ECGI), Law Working Paper N° 130/2009, April 2010, p.4.
[38] H. Kent Baker and R. Anderson, "An Overview of Corporate Governance", in "Corporate governance: a synthesis of theory, research and practice", 2010, p. 15.
[39] First published in 1999, the "Principles of Corporate Governance" of the OECD have been revised in 2004 and in 2015 by the OECD and the G20.
[40] OECD, "G20/OECD Principles of Corporate Governance", OECD Publishing, 2015.
[41] A. Gurria, "Note by the OECD Secretary General", G20 Finance Ministers and Central Bank Governors Meeting, 4-5 September 2015.

In our article, we have defined corporate governance as the mechanisms which have an impact inside a company/firm on the structure and on the relationship between the company's management, its board, its shareholders and other stakeholders, and whose aim is to provide an efficient structure to control and monitor effectively the company's management, objectives and performance. We will also broaden the scope of entities covered by corporate governance analysis to have a comprehensive view of the tokenization potential impacts on various kinds of corporate entities, including SMEs[42] and publicly traded companies.

ii. How could corporate governance be disrupted by DLT?

In this paper two perspectives are developed under which DLT could disrupt corporate governance. The first perspective is that the use of DLT could consist in "retrofitting"[43] and in strengthening the digitalization of companies' management and help solve issues already addressed for decades by professionals and academics.

For years there is an ongoing debate on how the development of technology would impact corporate governance and would further encourage the implication of shareholders and a more effective control of the management's activity, especially in the publicly traded companies. The possibility to use DLT for e-voting is relevant to this end[44]. Scholars Lafarre and Van Der Elst are describing that DLT can "offer smart solutions for classical inefficiencies in the corporate field," and with a "special attention to the restructuring of the old-fashioned and rigid Annual General Meeting of Shareholders" (the "**AGM**")[45]. According to the authors, distributed ledger technology can lower shareholder voting costs and the AGM organization costs for firms. Moreover, DLT has the potential to increase the speed of decision-making, facilitate a fast and efficient involvement of shareholders, and disintermediate and modify the role of proxy firms[46]. DLT could also help to address longstanding issues related to the transparency and accuracy of the identity verification of the shareholders[47], to minimize "disguised derivatives hedging, backdating and similar undesirable actions"[48], and be presented as a tool to strengthen shareholder participation.

One basic example is the possibility to provide virtual or hybrid – i.e. part physical and part virtual – shareholder meetings. A shift in that direction has started by amending companies' articles of associations to permit virtual-only or hybrid shareholders' general meetings, in order to provide more flexibility and lower cost for their organization, and to adapt to the internationalization and institutionalization of their shareholders' identity. A "virtual-only" shareholders' general meeting would be held without the need of determining a physical location and organized only by digital technology (e.g. a conference call or a dedicated and web-based interface/application), while a "hybrid" shareholders' meeting would provide to shareholders both physical and remote electronic accesses to the meeting. In a poll released by the association of general counsels of UK companies listed in FTSE 100[49], 62 per cent of their participating members indicated that their companies are planning to change their articles of associations to allow such meetings, and one third proposed to permit hybrid meetings.[50]

---

[42] The definition of Small and Medium-sized Enterprises (or "SMEs") by the European Union recommendation n°2003/361: http://ec.europa.eu/growth/smes/business-friendly-environment/sme-definition_fr

[43] Retrofitting is defined by Fenwick and Vermeulen as "adding digital solutions to older systems, models and organizations in the belief that this will "future proof" an existing approach and make it more efficient." Cf. M. Fenwick and E. Vermeulen, "Technology and Corporate Governance: Blockchain, Crypto and Artificial Intelligence", ECGI Working Paper No. 424/2018, November 2018, p. 13.

[44] P. Boucher, "What if blockchain technology revolutionized voting?", European Parliamentary Research Service, September 2016.

[45] A. Lafarre and C. Van Der Elst, "Blockchain Technology for Corporate Governance and Shareholder Activism", ECGI.com, Law Working Paper N° 390/2018, March 2018; A. Lafarre and C. Van der Elst, "Blockchain and the 21st century annual general meeting", European Company Law Journal 14, no. 4, 2017.

[46] J. Laster, "The Blockchain Plunger: Using Technology to Clean Up Proxy Plumbing and Take Back the Vote", Keynote Speech, Council of Institutional Investors, Chicago, 29 September 2016.

[47] D. Yermack, "Corporate Governance and Blockchains", Oxford Review of Finance, Volume 21, March 2017; M. Kahan and E. Rock, "The hanging chads of corporate voting", Georgetown Law Journal 96, 2008, pp. 1227-1281.

[48] V. Akgiray, "Blockchain Technology and Corporate Governance", Report for the OECD Corporate Governance Committee's roundtable discussion on blockchain technologies and possible implications for effective use and implementation of the G20/OECD Principles of Corporate Governance, 6 June 2018, p.24.

Several implications would arise to pass from a purely physical meeting of shareholders to a virtual or mixed meeting of shareholders. Virtual-only or mixed physical-digital meetings would represent a massive paradigm change, as it would allow shareholders to reduce transportation costs for shareholders' attendance to the meetings and facilitate the exchanges and Q&As with shareholders via electronic means. Nevertheless, some market associations have raised several drawbacks related to virtual-only meetings: the difficulty to provide sufficient technical certainty for the virtual access to the meeting (especially for companies with large numbers of shareholders), and the reputational risk (the perception that the management wants to avoid public physical meetings with shareholders). Meanwhile, it is almost certain that companies will develop the possibility for hybrid meetings in the years to come.[51]

This evolution towards the use of digital technology for the organization of AGMs is potentially a clear use case for the distributed ledger technology to provide "efficient and fair shareholders' meetings"[52], potentially streamlining the proxy voting process[53]. A private DLT, or a public DLT for greater transparency, could be created to register financial assets or records, or to allow shareholders using multiple signature (or multi-sig) technology[54] to address proposals to the management of the company, or the company's management to obtain shareholders' feedback on strategic matters, under pre-determined voting conditions (thresholds…), and in accordance with the applicable corporate regulations. The shareholders would be instantly notified of requests proposed either by other shareholders or by the management and could vote online – almost instantly – on these proposals. Majority rules, access rights, shareholder identification and legal requirements could be stored and encrypted in smart contracts and would be executed automatically during the voting process. These DLT shareholders' votes in real time could affect basic corporate governance issues such as the distribution of profits and liquidation surplus, information requests, etc...

b. <u>Legal and economic potential impacts of a new kind of corporate stakeholders: the token holders</u>

---

[49] The GC100, or the Association of General Counsel and Company Secretaries working in FTSE 100 Companies.
[50] D. Currie and N. Delaney, "Virtual shareholder meetings – stepping into Jimmy Choo's shoes or a matter of bad practice?", Reed Smith, 31 May 2018.
[51] Ibid.
[52] V. Akgiray, "Blockchain Technology and Corporate Governance", Report for the OECD Corporate Governance Committee's roundtable discussion on blockchain technologies and possible implications for effective use and implementation of the G20/OECD Principles of Corporate Governance, 6 June 2018, p.24; A. Lafarre and C. Van Der Elst, "Blockchain Technology for Corporate Governance and Shareholder Activism", ECGI.com, Law Working Paper N° 390/2018, March 2018; A. Lafarre and C. Van der Elst, "Blockchain and the 21st century annual general meeting", European Company Law Journal 14, no. 4, 2017.
[53] A. Irrera, "Nasdaq successfully completes blockchain test in Estonia", Reuters, 23 January 2017.
[54] A. Wright and P. de Filippi, "Decentralized Blockchain Technology and the Rise of Lex Cryptographia", SSRN, March 2015; V. Buterin, "Multisig: The Future of Bitcoin", Bitcoin Magazine, 12 March 2014.

In addition to the continuation of the digitalization of corporate governance, another impact could emerge due to the rights granted to new company stakeholders named crypto-asset holders (or token holders).

Crypto-assets are not a monolithic type of assets and each crypto-asset should be considered based on its specific features and functions. As every existing asset type can be - in theory at least - tokenized and as tokens could allocate rights to their holders specific to each issuance, the regulatory characterization of crypto-assets should be a clear point of attention for market participants. From a long-term perspective, with the development of token creations by listed companies, financial institutions and public entities, requests by major participants and regulators to categorize crypto-assets and to classify them along 'mainstream' token asset classes will have to be assessed, through the creation of groups of tokens representing similar asset types that could become susceptible to a certain degree of categorization.[55] But in the meantime, legal classification of crypto-assets, and the potential contradictions between national/regional regulations, will be a real challenge for businesses[56].

In this paragraph, the potential effects on corporate governance of security tokens, of utility tokens and, to a lesser extent, of cryptocurrency tokens will be examined. This paragraph will not detail precisely the differences between the three main types of crypto-assets (security, utility and cryptocurrency tokens), but will instead explain how the characterization by regulators of crypto-assets as such would have consequences on the governance of companies.

  i. The complex characterization of security tokens as securities

There is an intrinsic difficulty to classify several types of tokens as asset classes, but this classification issue appears to be in theory less problematic for security tokens. Since 2016, regulators from several countries have published reports and public positions arguing that the tokens which would be characterized as "securities" or "financial instruments" under their current rules would have to abide by the relevant local security laws. This is notably the case in the United States, in France[57], in Great Britain[58], in Switzerland[59], in Australia[60], in Canada[61] or in Singapore[62].

---

[55] Morgan Stanley, "Update: Bitcoin, Cryptocurrencies and Blockchain", research report, 31 October 2018; D. Bianchi, "Cryptocurrencies as an asset class? An empirical assessment", WBS Finance Group Research Paper, June 2018; J. Antos, "An Efficient-Markets Valuation Framework for Cryptoassets using Black-Scholes Option Theory", Medium, March 2018; L.W. Kong, E.C. Laurenson, A.C. Scheibe, D.L. Taub, L. Tessler, V. Van Tassel Richards, "Five Regulatory Implications for Blockchain Tokens as an "Asset Class", McDermott Will & Emery, 10 January 2018; C. Burniske and J. Tatar, "Cryptoassets", 2017.
[56] I. Barsan, "Legal Challenges of Initial Coin Offerings (ICO)", Revue Trimestrielle de Droit Financier (RTDF), n°3, 2017, pp. 54-65.
[57] French financial regulator AMF, "Summary of replies to the public consultation on Initial Coin Offerings (ICOs) and update on the UNICORN Programme", 22 February 2018.
[58] Financial Conduct Authority, English HM Treasury and Bank of England, "Cryptoassets Taskforce: final report", October 2018.
[59] Directorate General for Economic Development, Research and Innovation (DG DERI) of the State of Geneva, "Guide: Initial Coin Offerings (ICOs) in the Canton of Geneva", 28 May 2018; FINMA, "Guidelines for enquiries regarding the regulatory framework for initial coin offerings (ICOs)", 16 February 2018.
[60] Australian Securities & Investments Commission (ASIC), "Initial coin offerings", Info 225, September 2017.
[61] Canada Securities Administrators, "Cryptocurrency Offerings", CSA Staff Notice 46-307, 24 August 2017; Ontario Securities Commission (OSC), "OSC highlights potential securities law requirements for businesses using distributed ledger technologies", Press release, 8 March 2017.
[62] "Guide to Digital Token Offerings" published by the Singapore financial regulator Monetary Authority of Singapore (MAS) in 14 November 2017; MAS, "MAS clarifies regulatory position on the offer of digital tokens in Singapore", Press release, 1 August 2017.

There is a growing consensus among regulators, market participants and law firms from OECD countries to assume that security tokens are equivalent to securities[63]. Under such assumption, it could appear easier to define from a regulatory perspective security tokens than utility tokens and therefore its relevant applicable legal and operational framework would be easily assessed. The reality is much more complex, at least at this stage given the present knowledge of the DLT and crypto-asset development by market participants. Major market participants are awaiting clarification from legislators and regulators, even the US where almost all the existing tokens, with the exception of Bitcoin and perhaps also Ether, are characterized as securities[64], that security tokens are fully assimilated to securities, from a legal, tax, prudential and accounting[65] perspective. And even if this assimilation between security tokens and securities is provided on a national basis, it would nevertheless require a global coordination, as securities such as equity or debt have different legal status and regime in different jurisdictions.

At the corporate governance level, the main issue for security token holders is to know whether these investors could truly be legally considered and operationally[66] treated as 'traditional' security holders, and how they can exercise the rights granted to them. The reality is that it remains to be clarified, according to the project envisioned and to regulatory stabilization, that the purchase of security crypto-assets (during ICOs, on crypto-exchange platforms, by over-the-counter transactions, or else) could have similar qualifications and effects than the purchase of 'traditional' securities such as equity or debt instruments.

This is a critical issue for holders of security tokens which possess some characteristics similar to equities ("**equity tokens**"). Equity tokens are, with debt tokens, one of the sub-categories of security tokens. In most cases of the current existing ICOs, equity tokens are not granting all the rights associated to equity instruments (ownership and control rights), or "have comparable features to investments […] but are structured in such a way that they fall outside the regulatory perimeter (either intentionally or not)."[67] For example, tokens could avoid granting voting rights to investors, rights to dividends and/or to liquidation surplus, or allowing them to submit the inclusion of a draft resolution to shareholder meetings[68]. If security token holders do not have the same rights than security holders, and notably in relation to shares, could they truly be considered as shareholders, and equity tokens be assimilated to 'traditional' shares?

---

[63] This consensus is yet to be confirmed, e.g. the final report of English regulators "Cryptoassets Taskforce from October 2018: "While security tokens fall within the current regulatory perimeter and it is the responsibility of firms to determine whether their activities require authorization, the Taskforce recognizes that the complexity and opacity of many cryptoassets means it is difficult to determine whether they qualify as security tokens."

[64] J. Rohr and A. Wright, "Blockchain-Based Token Sales, Initial Coin Offerings, and the Democratization of Public Capital Markets", Cardozo Legal Studies Research Paper No. 527, 5 October 2017.

[65] The French accounting national authority (the "Autorité des Normes Comptables") is supposed to publish a position on tokens' accounting treatment (French GAAP) by the end of 2018, but the discussions on international standards for utility tokens (and not yet for security tokens) will be only started from the first quarter of 2019 by the "International Accounting Standards Board" ("IASB").

[66] From an operational perspective, in the first generation of security tokens offerings, all the token management was provided off-chain or with a mixed off-chain and on-chain in order to assess the distributed ledger technology, but as the security token market will grow the market participants will try to develop as much as possible on-chain token management in order to reduce back offices and middle offices costs and to prevent a double booking (physical and digital) of the operations.

[67] Financial Conduct Authority, English HM Treasury and Bank of England, "Cryptoassets Taskforce: final report", October 2018, p. 20.

[68] French financial regulator AMF, "Summary of replies to the public consultation on Initial Coin Offerings (ICOs) and update on the UNICORN Programme", 22 February 2018.

Equity tokens are often described as one of the most promising crypto-asset classes[69]. This type of crypto-assets could have several qualities and value in themselves which could interest corporate executives. Crypto-assets, and therefore equity tokens, could be in theory traded all year long and without any boundary or geographical time limitations on crypto-asset exchanges, contrary to traditional corporate shares and financial instruments[70]. Besides, vesting periods as well as specificities related to the equity tokens can be easily introduced in the smart contract related to those crypto-assets.

In addition, equity tokens could have also the potential to modify the relationship between company founders and investors. There has been a trend for tech companies (e.g. Snap[71], Facebook, Google, LinkedIn, Groupon) to alter the "one share, one vote" principle and to reduce or avoid granting voting rights for non-founding investors by creating different classes of shares for the different kinds of shareholders (founders, early investors, late investors…), and by issuing multiple voting shares as well as non-voting shares[72]. This trend, which has been explained frequently by the will to protect the ability for the company founders and innovators to keep a clear control of the company, to modify the influence of institutional investors and shareholder activists, and to create value with a long-term perspective, could find its continuity in the development of ICOs and security tokens created by blockchain tech companies. Crypto-assets have been created by many start-ups at an early stage where founders have wanted to maintain a clear control of the shareholding and a clear majority for the main voting decisions by the newly-created firm.

While it could be a useful tool for entrepreneurs to avoid overly direct pressure by shareholder activists on the company's profitability at an early stage of the company, issuing equity tokens without granting voting rights, sometimes without a financial compensation[73], or holding their own equity tokens in treasury[74] could represent a challenge for the company in the long term, beyond legal issues[75], at a reputational level and could lead to shareholders' reaction through public positions, the kind of situation Facebook has already met in September 2017[76]. Equity token holding will evolve, and it could push companies' management team to provide innovative ways to involve equity token holders in the company decision process (e.g. providing regular virtual consultations and ad hoc security token holders' general assemblies, issuing diminished rather than non-voting rights, creating holding periods for preferred share classes designed for long-term crypto-investors…).

---

[69] H. Marks, "The future of US securities will be tokenized", Medium, 22 May 2018.

[70] Provided that, without a clear international political agreement or international treaty on token status, there is a legal uncertainty as to how a given jurisdiction may qualify an equity token issued in another jurisdiction, as well as recharacterization and liability risks for the issuer and the investors. Legal advice is recommended.

[71] "Snap Inc.'s IPO [on March 2, 2017], featuring public shares with no voting rights, appears to be the first no-vote listing at IPO on a US exchange since the New York Stock Exchange (NYSE) in 1940 generally barred multi-class common stock structures with differential voting rights", in: K. Bertsch, "Snap and the Rise of No-Vote Common Shares", Harvard Law School Forum on Corporate Governance and Financial Regulation, 26 May 2017; Les Echos, "Snap innove en émettant des actions sans droit de vote à l'occasion de son IPO", 2 March 2017.

[72] S.C.Y.Wong, "Rethinking 'One Share, One Vote'", Harvard Business Review, 29 January 2013.

[73] An "airdrop" consists in giving tokens to early adopters without any financial contribution in order to raise the popularity of the token or the project. See: K. Bheemaiah and A. Collomb, "Cryptoasset valuation: Identifying the variables of analysis", Louis Bachelier Institute, Working Report v1.0, 19 October 2018, p. 16.

[74] Contrary to the so-called "treasury stocks" or "treasury shares", issuers can hold their own equity tokens as soon as the issuance itself, and not through buybacks of equity tokens. Self-holding of tokens by their own issuers is a clear issue that should be tackled in the future in corporate governance studies related to DLT.

[75] In French law, there is a clear limitation on shares without voting rights (which are not considered as "ordinary shares" but as preference shares or "actions de preference") issued by joint-stock companies, which is half of the share capital for non-listed companies and a quarter of the share capital for listed ones, as provided in article L.228-11 of the French commerce code. Several countries such as Belgium or Sweden have enacted regulations to limit the disparities between voting rights attached to shares.

[76] Les Echos, "Actions sans droite de vote : les investisseurs gagnent une bataille", 25 September 2017.

A second question is that, even though security tokens holders are considered as 'traditional' security holders, how could they exercise their rights? For example, if cryptographic private keys of equity tokens are stolen by an important investor, how would be assessed such loss by the investor, by its custodian, and by the company itself? And if a company issues security tokens, would it be possible to tokenize all the prior existing securities? As of today, regulated stock exchanges (e.g. New York Stock Exchange, NASDAQ exchanges, Euronext) are not yet able to list securities-tokens ICOs nor to assess security tokens instead of 'traditional' company shares. And even regulations from G7 countries have not already assessed the various implications to tokenize existing listed securities[77]. One additional layer of difficulty is the potential difference between market capitalization and market tokenization, especially for listed companies. Even if an equity token is deemed equivalent to a share, will it have the same market value than a 'traditional' share? Given the current questions related to the volatility, to the valuation methodologies and to the liquidity of tokens[78], and due to the fact that equity tokens are not yet traded on 'traditional' stock exchanges but currently only on crypto-exchanges, it remains to be verified whether an equity token A equivalent to a non-tokenized share B would have the same value than this share B[79]. Stock market platforms willing to list equity tokens and token-only exchanges will have to handle market abuse concerns[80] and to provide fair and effective price discovery and cross-listing transparent information for these new assets.

Equity tokens also pose the issue of the identification of shareholders. Until now, the acquisition of almost every existing token is based on pseudonymity. While the owners of tokens are known by a pseudonym, the precise knowledge of shareholders' identities is an absolute necessity for a company[81]. Many blockchain entrepreneurs are against the concept of allowing the exact identification of the token holders, but as the crypto-asset market grows and develops to a massive adoption by major market participants, market practices will emerge and provide ways to provide investor whitelisting/validation processes and to identify precisely the holders of the security tokens[82]. Without the possibility of verifying shareholders' identities, mergers and acquisitions of companies financed by crypto-assets and the use of DLT, as well as class actions, and even the dissolution of the company which has issued the equity tokens, would become much more difficult, as it would be technologically difficult to contact all the shareholders and security token holders. Several systems have been created to develop the identification of security holders[83], and therefore to "reduce the opportunity for rent-seeking or corrupt behavior"[84]. Even though the possibility of human or technological errors in transfer of equity tokens remains[85], the transparency provided by an equity token registration on a public DLT could permit anyone (investors, market makers…) to observe in real time managers' and other shareholders' ownership positions and trading activity[86].

---

[77] For example, in France, a new regulation has allowed the registration of securities' issuances and transfers on a distributed ledger, but this interesting initiative has focused for the moment only on non-listed shares, debt instruments and units of OPCs, for experimental reasons. The same limitation to unlisted securities has also been provided in a Senate Bill chaptered in September 2018 in the US State of California. Sources: French Law n° 2016-1691 of 9 December 2016 "relative à la transparence, à la lutte contre la corruption et à la modernisation de la vie économique" (the so-called "Sapin II law") ; Ordonnance n°2017-1674 of 8 December 2017 "relative à l'utilisation d'un dispositif d'enregistrement électronique partagé pour la représentation et la transmission de titres financiers"; California Senate Bill No. 838, chaptered by the California Secretary of State on 28 September 2018 in Chapter 889 of Statutes of 2018 ; Reuters, "France to allow blockchain for trading unlisted securities", 8 December 2017; K. Lachgar and J. Sutour, "(R)évolution blockchain pour les titres non cotés français : enjeux et perspectives autour de la consultation publique de la Direction Générale du Trésor", LEXplicite.fr, 9 June 2017.

[78] "Due to their potential to reduce costs and shorten the time required for executing and settling securities trades, blockchains offer the possibility of significant improvements in liquidity, whether they are used as the main platform for share registration and exchange, or alternatively, whether they are introduced by stock markets in a more limited way to streamline the post-trade clearing and settlement process.", in D. Yermack, "Corporate Governance and Blockchains", Oxford Review of Finance, Volume 21, March 2017, pp. 7-31.

[79] "There is a large amount of assumptions being made, the key one being that a model or a formula that is used for stock valuation can be applied to a new asset class which has very different properties", and "it is important to highlight that fundamental analysis practices that are applied to stock analysis [i.e. classical due diligence practices on the project and the project team] still apply to tokens", but "there can be no one-size-fits-all [token] evaluation methodology as seen with stocks". "While stock evaluation is largely made up of financial variables and ratios, tokens are fully digital entities that exist on a network plane. Thus the kinds of variables that need to be analyzed are not just financial but technical as well, especially when analyzing smart contracts", in K. Bheemaiah and A. Collomb, "Cryptoasset valuation: Identifying the variables of analysis", Louis Bachelier Institute, Working Report v1.0, 19 October 2018, pages 6 and 20.

[80] R. Keidar and S. Blemus, "Cryptocurrencies and Market Abuse Risks: It's Time for Self-Regulation", SSRN, 25 February 2018.

[81] Or for shareholders willing to prepare a pooling agreement.

    ii. The difficulties for utility and crypto-currency tokens holders to be recognized as corporate stakeholders

While there are few documented studies on the matter, utility tokens and cryptocurrency tokens could also pose important questions related to corporate law and to corporate management. It is a common mistake to think that utility tokens will not have any regulatory or economic implications in the future. In start-up financing, many funders have been inclined to create utility tokens in order to raise funds without granting to investors economic or political rights nor having any substantial fiduciary duty to the investors[87]. But many utility tokens are already exchanged on crypto-currencies exchanges, and therefore have an economic value, at least on the secondary market.

---

[82] Anti-money laundering (AML) and know-your-customer (KYC) processes will progressively be standardized and provided by market participants, such as crypto-exchanges, custodian wallet providers, crypto-brokers and banks, willing to develop the crypto-asset digital financial markets. Cf.: US Department of the Treasury, Financial Crimes Enforcement Network, "Guidance on the Application of FinCEN's Regulations to Persons Administering, Exchanging or Using Virtual Currencies", FIN-2013-G001, 18 March 2013; Directive (EU) 2018/843 of the European Parliament and of the Council of 30 May 2018 amending Directive (EU) 2015/849 on the prevention of the use of the financial system for the purposes of money laundering or terrorist financing; D. Holman, B. Stettner, "Anti-Money Laundering Regulation of Cryptocurrency: US and Global Approaches", Allen & Overy LLP, published in "Anti-Money Laundering Laws and Regulations 2018", ICLG, July 2018; D. Siegel and others, "The Equity Token Project", report to be published in 2019.
[83] The role of the custodian of crypto-asset private cryptographic keys will be essential in the years to come. Defining its precise role and its specificity from the ones of the existing custody services will be an essential debate of the future crypto-asset regulations. The Directive (EU) 2018/843 defines "custodian wallet provider" as "an entity that provides services to safeguard private cryptographic keys on behalf of its customers, to hold, store and transfer virtual currencies." In France, this debate has occurred during the discussions on the adoption of the "PACTE" draft bill, and on its provisions related to crypto-assets service providers (in its Article 26 bis A).
[84] In the long term, when DLT will be broadly used by companies, and notably public ones, regulators might be interested to provide mandatory disclosures of public keys by equity token holders. See: D. Yermack, "Corporate Governance and Blockchains", Oxford Review of Finance, Volume 21, March 2017, pp. 7-31; K. Malinova and A. Park, "Market design with blockchain technology", SSRN, 30 May 2016.
[85] A. Tinianow, "Tokenized securities are not secured by Delaware Blockchain Amendments", 4 July 2018.
[86] D. Yermack, "Corporate Governance and Blockchains", Oxford Review of Finance, Volume 21, March 2017. See also: A. Edmans, "Blockholders and corporate governance", Annual Review of Financial Economics, Vol. 6, 2014, pp. 23-50; L. Bebchuk and R. Jackson, "The law and economics of blockholder disclosure", Harvard Business Law Review, 2012, pp. 39-60.
[87] K. Bheemaiah and A. Collomb, "Cryptoasset valuation – Identifying the variables of analysis", Louis Bachelier Institute, Working Report v1.0, October 2018, p.16.

Utility tokens, as well as cryptocurrencies, as they can be exchanged on the secondary market, could represent an important shift in the way we consider corporate stakeholders beyond the sole shareholder issue. While holders of these utility tokens or cryptocurrency tokens will not have voting rights during AGMs, reserved to equity token holders, it could be imagined that the market value of these tokens on crypto exchanges would represent an important role in exerting pressure on the company management and for these stakeholders to have an indirect impact on the company's decisions. The day will come when a renowned listed corporation, such as Amazon[88], Carrefour[89], Facebook[90] or Société Générale[91], will issue a utility token whose value can be exchanged on the secondary market and it will be possible for the utility token holders, and notably the consumers of a company, to develop a direct dialogue with the corporation and to send requests to the company management. In the long term, companies could have to rethink their business models[92], the role of these stakeholders, and create ad hoc assemblies, specific committees or other innovative involvement modes in order to develop the interactions with these new company stakeholders. After the creation of assembly meetings of shareholders and bondholders, the day may come when it is necessary to organize general assemblies of utility token holders[93]. In order to avoid a recharacterization of the utility tokens into security tokens, these general meetings would not consist in votes on the company management, but on a new kind of communication with the company's stakeholders. Issuing utility tokens, or even stable tokens[94], could be a way for companies to attract and reward early adopters of the tokens[95], to provide additional rights to their consumers, shareholders, suppliers, investors and/or employees, and an incentive to promote a deeper commitment to the firm from them. Informal assemblies of utility token holders, organized with only noncompulsory consultations, could be used to develop and encourage interaction, communication and disclosures of reportings with the vast company's ecosystem, not through a voting process but through more consultation with the various stakeholders in the ecosystem of the company.

Beyond the focus on shareholders, in the context of a movement towards a more digitalized, platformized and decentralized economy which is an intrinsic part of DLT, a new role could be attributed with such technology to the various types of corporate stakeholders[96]. In an expert report on the future of corporations for the French government published in 2018, one of the key proposals is the creation of stakeholders' committees in companies[97]. According to Dondero, the current company law "organizes the relationships between partners/shareholders, between them and the directors and the other corporate bodies, and to some extent, it apprehends the situation of employees and of the CSR[98]."[99] Some proposals in France were to extend the "affectio societatis" to a wider scope of people and stakeholders of the company, by recognizing that the role of companies in society evolves to roles close to the ones of states and public institutions. The creation of crypto-assets, and even more of crypto-currencies, is just the next step towards this evolution. As companies may have the ability, and the right, to create their own currency, the expansion of their obligations on an international regulatory level could appear as a necessity.

---

[88] E. Kim, "Amazon just bought three domain names related to cryptocurrency", CNBC, 1 November 2017.
[89] Reuters, "Retailer Carrefour using blockchain to improve checks on food products", 6 March 2018.
[90] A. Cuthbertson, "Is Facebook about to launch its own cryptocurrency?", The Independent, 9 May 2018.
[91] Blockchain is an "emerging technology we are exploring, and which is expected to stabilize", speech by Philippe Heim, Deputy Chief Executive Officer of Société Générale, during the Blockchain conference organized by the Sapiens Institute at Sciences Po on 25 September 2018.
[92] For example, by focusing more about developing micropayments systems automated and implemented by smart contracts, instead of relying on "advertising-based revenue models". Cf. A. Wright and P. de Filippi, "Decentralized Blockchain Technology and the Rise of Lex Cryptographia", SSRN, March 2015, p.30.
[93] Even though existing ICO whitepapers rarely provide assembly meetings of token holders, with the exception of some security token issuances (e.g.: Alethena Token Specifications published on 14 May 2018).
[94] A stable token (or stable coin) is a crypto-asset whose value is pegged to an existing asset (mostly fiat currencies such as US Dollars and Euros, but also sometimes gold), and which can be collateralized by this existing asset. Stable tokens have been mostly used in 2018 to lessen the volatility related to crypto-assets and to provide a certain financial stability solution at the crossroads between crypto-asset markets and fiat currencies markets. It is too soon to report any existing regulation specifically dedicated to this new kind of crypto-assets.
[95] C. Catalini and J. Gans, "Some Simple Economics of the Blockchain", MIT Sloan Research Paper No. 5191-16, 21 September 2017, p 17.
[96] M. Fenwick and E. Vermeulen, "Technology and Corporate Governance: Blockchain, Crypto and Artificial Intelligence", ECGI Working Paper No. 424/2018, November 2018, pp.16-19.

### III. The emergence of distributed governance

After an analysis of the potential impacts of DLT, ICOs and crypto-assets on the governance of existing companies based on centralized governance, Part III addresses the current debates on the future of firms and their potential disruption caused by a new type of governance, named thereafter the "**distributed governance**", which could substantially affect the way organizations are currently governed.

The aggregation of several smart contracts could encrypt, provide and implement governance rules. There is therefore an increasing debate on the possibility that the use of smart contracts could substantially impact corporate governance, improve its efficiency, and beyond that the combination of smart contracts with DLT networks could foster the creation of new kinds of organizations[100]. The failure of the start-up "The DAO", announced in 2016 as being a symbol of a new type of human organization, i.e. decentralized autonomous organizations, will be investigated in depth in this Part III, as well as the description of what could constitute a decentralized/distributed form of corporate governance based on peer-to-peer cooperation and on consensus (or, more accurately, majority-based) automated decision making processes.

a. The lessons from the "The DAO" case
  i. What is the "The DAO" case?

The decentralized ledger technology has been thought as a way to rely less on centralized management structures. In 2016, a project named "The DAO" has been developed by computer scientists such as Christoph Jentzsch and presented to the public as representing the creation of a new kind of organization, i.e. the "decentralized autonomous organization" ("**DAO**"). "The DAO" governance was not based on a traditional form of management (a CEO, a Chairman and a board of directors) but rather based on an investor-based management where the corporate decisions to invest in projects were directly decided and controlled by crypto-asset holders' online voting processes. "The DAO" project has failed, but the explanation of its failure is necessary to study further proposals related to the emergence of DAOs as a new type of organization.

---

[97] N. Notat and J.-D. Senard, "L'entreprise, objet d'intérêt collectif – Rapport aux Ministres de la Transition écologique et solidaire, de la Justice, de l'Economie et des Finances, et du Travail", 9 March 2018.
[98] Corporate Social Responsibility or "CSR".
[99] B. Dondero, "La raison d'être des entreprises (rapport Notat-Senard)", 10 March 2018: https://brunodondero.com/2018/03/10/la-raison-detre-des-entreprises-rapport-notat-senard/
[100] U. Chohan, "The Decentralized Autonomous Organization and Governance Issues", SSRN, Discussion Paper, December 2017; D. Yermack, "Corporate Governance and Blockchains", Oxford Review of Finance, Volume 21, Issue 1, March 2017, pp. 7-31; P. Paech, "The Governance of Blockchain Financial Networks", LSE Legal Studies Working Paper No. 16/2017, 16 December 2016; M. Raskin, "The Law and Legality of Smart Contracts", Georgetown Law Technology Review, Volume 1, 25 September 2016, p. 336.

Described in its white paper as being the "first implementation of Decentralized Autonomous Organization (DAO) code to automate organizational governance and decision making"[101], the project "The DAO" was a for-profit, decentralized, crowdfunded, presented as a direct management (or direct-democracy) organization and an investment platform running on the Ethereum platform "that would create and hold corpus of assets through the sale of DAO tokens to investors, which assets would then be used to fund 'projects'"[102]. Anyone could invest ether in "The DAO" project to acquire "The DAO" crypto-assets and would then be allowed to vote on the investing projects proposed by "The DAO" funders and have a "share in the anticipated earnings"[103] from the projects funded as a return on investments, as well as re-selling "The DAO" crypto-assets on the secondary market.

On June 17, 2016, The DAO had raised more than 11 million ethers, or in value about 143 million US dollars at the time, through their ICO from close to 20 000 investors. But in just a few seconds the organization was robbed of 3,6 million ethers, or about 50 million US dollars. The hackers, who used a bug in the "The DAO" code, could not recover the money which was blocked for a month. This delay has allowed "The DAO" stakeholders to vote online and to put in place measures during this time to recover the money, by deciding to use the hard fork process. This event has called into question all the ideas around the blockchain: digital innovations, open source, protection of stakeholders, liability and indemnity issues, encrypted operations guarantee, etc.

The "The DAO" white paper provided that a DAO "can be used by individuals working together collaboratively outside of a traditional corporate form. It can also be used by a registered corporate entity to automate formal governance rules contained in corporate bylaws or imposed by law." The white paper proposed that a DAO "would use smart contracts to attempt to solve governance issues it described as inherent in traditional corporations". A DAO "purportedly would supplant traditional mechanisms of corporate governance and management with a blockchain such that contractual terms are "formalized, automated and enforced using software"". As for the legal status of DAOs, the white paper posited that it "remains the subject of active and vigorous debate and discussion. Not everyone shares the same definition. Some have said they are autonomous code and can operate independently of legal systems; others have said that they must be owned or operate[d] by humans or human created entities. There will be many use cases, and the DAO code will develop over time. Ultimately, how a DAO functions and its legal status will depend on many factors, including how DAO code is used, where it is used, and who uses it."[104]

---

[101] Christoph Jentzsch, "Decentralized Autonomous Organization to Automate Governance", Final Draft – Under Review, 2016.
[102] Securities and Exchange Commission, "Report of Investigation pursuant to Section 21(a) of the Securities Exchange Act of 1934: The DAO", Release No. 81207, 25 July 2017.
[103] Ibid.
[104] Ibid., page 3.

In its investigative report on the issuance by "The DAO" published on 25 July 2017, the US federal financial regulator – i.e. the Securities and Exchange Commission – has considered that the token issued by the "The DAO" project was a security and that its creators should have complied with US federal financial regulations.

ii. Corporate governance and the definition of DAO as a nexus of computer code contracts

While the "The DAO" project was a short-lived experiment, there is an increasing interest by DLT entrepreneurs on the potential use of decentralized autonomous organizations (DAO)[105]. The most famous entrepreneurs of the DLT ecosystem have tried to define what a DAO is. A definition of DAOs has been proposed by Ethereum founder Vitalik Buterin as "a virtual entity that has a certain set of members or shareholders which, perhaps with a 67 % majority, have the right to spend the entity's funds and modify its code". It is a "long-term form of smart contract that contain the assets and encode the bylaws of an entire organization".[106] For Jentzsch, DAOs are "organizations in which (1) participants maintain direct real-time control of contributed funds and (2) governance rules are formalized, automated and enforced using software."[107]

From a technology perspective, a DAO is "a smart contract taking the form of organization of an undertaking by a group of people (and may be open to new members)"[108] whose primary purpose is to provide services and functions in the digital space. From a legal point of view, we define DAO as an organization created by one or more people agreeing to fulfill a common purpose by encoding rules about its governance, functions and/or operations in computer programs (the so-called 'smart contracts'), which is intrinsically not operating in a specific national state nor identified with any specified jurisdiction. DAOs can therefore carry out activities close to traditional existing institutions such as foundations, associations or companies, and have an impact on the way those entities are governed.

In a DAO, consensus (or, more accurately, majority-based) decisions by a crowd of token holders are supposed to replace individuals as managers. Through DAOs, computer code and smart contracts can replicate some aspects of corporate governance and allow parties to enjoy the benefits of corporate structures while maintaining the flexibility of informal online groups, for instance in facilitating the conditions of a digital voting in real time and in a transparent, auditable and tamper-proof manner without human intervention[109]. Such hypothesis where token holders are directly managing the organization without any management team has not been contemplated by traditional corporate governance rules. Indeed, according to the OECD Corporate Governance rules, "shareholders are not expected to assume responsibility for managing corporate activities," and that "the responsibility for corporate strategy and operations is typically placed in the hands of the board and a management team that is selected, motivated and, when necessary, replaced by the board."[110]

---

[105] P. Bloch, "Decentralized Autonomous Organization (DAO) et Théorie de l'Agence", LesEchos.fr, 31 May 2016.
[106] V. Buterin, "A next generation smart contract & decentralized application platform", Ethereum White Paper, 2014; S. Blemus, "Law and Blockchain: A Legal Perspective on Regulatory Trends Worldwide", Revue Trimestrielle de Droit Financier (Corporate Finance and Capital Markets Law Review), RTDF n°4-2017, December 2017.
[107] Christoph Jentzsch, "Decentralized Autonomous Organization to Automate Governance", Final Draft – Under Review, 2016.
[108] Wardyński & Partners, "Blockchain, smart contracts and DAO from a legal perspective", Oxford Business Law Blog, 12 January 2017.
[109] H. de Vauplane, "Blockchain et corporate gouvernance: stigmergie ou holacratie ?", Revue Banque n°821, 30 May 2018.

Even though DAOs could not be assimilated to a corporation legally recognized and incorporated as such under a specific jurisdiction[111], DAOs can nevertheless be considered as a firm. From the economic standpoint, DAO's construction as a network of smart contracts could be compared to the "nexus of contracts" described by Jensen, Meckling and Fama as the fundamental definition of firms[112]. According to Jensen and Meckling, a "private corporation or firm is simply one firm of legal fiction which serves as a nexus for contracting relationships and which is also characterized by the existence of divisible residual claims on the assets and cash flows of the organization which can generally be sold without permission of the other contracting individuals." A DAO is a nexus of contracts between the various holders of its tokens, and these crypto-asset holders could vote by a majority-vote on the investment allocations and could have a claim on the assets and the profits of the DAO. Computer codes could fit as "contracts" in the definition provided above by Jensen and Meckling, as their notion of "contract" is different from its accepted legal sense and is understood as "any mechanism in which ownership rights of property are created, modified or transferred"[113]. It remains to be seen whether countries will recognize a legal status for DAOs, providing what Jensen and Meckling refer to as "legal fiction", i.e. "the artificial construct under the law which allows certain organizations to be treated as individuals."

DAOs are based on the same idea provided by Jensen and Meckling that "contractual relations are the essence of the firm, not only with employees but with suppliers, customers, creditors, etc." But many issues remain to recognize DAOs as existing legal firms/corporations. Many DAO projects have been created by using the business model of a capitalistic stock company, but these DAOs have been created and their decisions enforced only on the Blockchain technology and not in existing legal jurisdictions. Besides, in theory, "DAOs are not governed by principal-agent relationships, since they do not have shareholders or managers"[114]. In order to provide a legal interim solution for DAOs which cannot yet be considered as a company or a legal entity (i.e. foundation, association…) recognized by a national country, some legal advisors have proposed to DAO founders to create and register a legal entity in a 'physical' jurisdiction which would represent the DAO in a given national country[115]. This entity would forge links between the digital and physical world and would permit to identify a specific jurisdiction, the applicable law, a recognized national legal status and to enforce DAO decisions before a court. In the long term, DAOs could only be effective if an international treaty concluded between sovereign nations would recognize a digital or distributed jurisdiction whose jurisdiction would be international and under a governance agreed by countries. Would the DAO as a legal entity only be recognized under a digital or distributed jurisdiction to be created on an international level? It is possible that there would be as many types of DAOs than there are types of tokens. But from a long-term perspective, the status of this smart contract-based form of organization called DAO will have to be further examined, notably on the consensus decision-making process.

---

[110] OECD, "OECD Principles of Corporate Governance", OECD Publishing, 2004, p.32; OECD, "G20/OECD Principles of Corporate Governance", OECD Publishing, 2015, p.18.
[111] Given that no country has yet recognized a legal status for this type of organizations.
[112] M. Jensen & W. Meckling, "Theory of the Firm: Managerial Behavior, Agency Costs and Ownership Structure", Journal of Financial Economics, Vol.3, Issue 4, pp. 305-360, October 1976; E. Fama & M. Jensen, "Separation of Ownership and Control", Journal of Law and Economics, Vol.26, No 2, pp. 301-325, June 1983; A. Wright and P. de Filippi, "Decentralized Blockchain Technology and the Rise of Lex Cryptographia", SSRN, March 2015, p.15.
[113] I. Tchotourian, "Théories contractuelles de la firme : Théorie du nœud de contrats et théorie de l'agence", http://droit-des-affaires.blogspot.com/2007/02/thories-contractuelles-de-la-firme_5992.html, February 2007.
[114] Y-Y. Hsieh, "The Rise of Decentralized Autonomous Organizations: Coordination and Growth within Cryptocurrencies", Thesis, University of Western Ontario, June 2018, p.106.
[115] Wardyński & Partners, "Blockchain, smart contracts and DAO from a legal perspective", Oxford Business Law Blog, 12 January 2017.

b. Distributed organization models

The potential of the existing financial assets' tokenization process could have substantial effects on corporate governance. Beyond the questions on DAOs, the trend goes from the dematerialization of assets to the disruption of the governance. We will analyze in the last part of this paper the new governance models for corporates based on a distributed and crowd-sourced decision-making process.

   i. Distributed governance and new models of corporate governance

The use of DLT technology, crypto-assets and smart contracts has allowed many innovators to think about new models of corporate governance. Developing consensus mechanisms for corporate decisions could alter the fundamentals of corporate governance, such as the firm theory[116], the agency theory and the relationship between agents and principals[117], beyond the traditional centralized and hierarchical governance structure of firms. Given that the word "consensus" frequently used in DLT literature refers to a "majority rule vote" for corporate governance matters, we will use the "consensus" wording to mean "majority rule vote" in the next paragraphs.

According to Jensen and Meckling, agency theory is based on "a contract under which one or more persons (the principal(s)) engage another person (the agent) to perform some service on their behalf which involves delegating some decision-making authority to the agent."[118] Distributed and consensus mechanisms are described by DLT theoreticians and by some academics as an instrument to solve several principal-agent issues and to foster the alignment of interests between the agents (i.e. the management) and the principals (i.e. the shareholders)[119]. The agency theory emphasizes potential risks related to the imperfect delegation of power by the principal(s) to the agent, notably due, according to Williamson, to the existence of information asymmetries and to the potential opportunistic behavior of the agent to the detriment of the principal[120]. The use of DLT and smart contracts could potentially provide a full and constant transparency and verifiability of the data available to every participant of the DLT - and notably minority stakeholders[121] - to control more effectively the information provided and potential opportunistic decisions of the corporate management[122]. In this way of thinking, the replacement of trust in a disruptive technology management instead of trust in a human management team would be a strong incentive to minimize agency costs.

---

[116] R. Coase, "The Nature of Firm", Economica, Vol. 4, No. 16, 1937.
[117] M. Jensen and W. Meckling, "Theory of the firm: Managerial behavior, agency costs and ownership structure", Journal of Financial Economics, October 1976, pp. 305-360.
[118] Ibid., p. 308.
[119] M. Fenwick and E. Vermeulen, "Technology and Corporate Governance: Blockchain, Crypto and Artificial Intelligence", ECGI Working Paper No. 424/2018, November 2018; P. de Filippi and A. Wright, "Blockchain and the Law: The Rule of Code", Harvard University Press, 2018; C. Reyes, N. Packin and B. Edwards, "Distributed Governance", William & Mary Law Review Online, Volume 59, 2017; M. Molines, "Comment la blockchain va changer la gouvernance des entreprises", TheConversation.com, 17 October 2017; A. Wright and P. de Filippi, "Decentralized Blockchain Technology and the Rise of Lex Cryptographia", SSRN, March 2015, p.15.
[120] O. Williamson, "Markets and Hierarchies – Analysis and Antitrust Implications: a Study in the Economics of International Organization", Free Press, 1975.
[121] The cost of accessing information for minority stakeholders would be potentially reduced. Cf. S. Burger, "Blockchain Technology Can Improve Multistakeholder Process Integrity", Engineering News, 12 August 2016.

This perspective towards distributed governance could have implications in various activities. One example is the Klerios start-up which has developed a decentralized justice governance based on ancient Greek legal democracy (the Kleroterion), by providing the resolution of disputes through the selection of jurors chosen randomly or based on a reputational basis[123]. In this project, game theory would apply, in a form close to the prisoner's dilemma: jurors would not deliberate, just vote without communicating between each other and without knowing the vote of the other jurors.

The modalities towards distributed governance is subject to important expert debates. Voter participation is a clear focus of corporate governance, and consensus mechanisms represent a real challenge for corporate governance. Will the governance of DLT projects be fully integrated on-chain, or part on-chain and part off-chain? An on-chain governance model would consist in providing all the governance through online voting by token holders on the DLT, similar to a modern direct democracy[124]. The major issue of this equivalency between a distributed ledger users and citizens in a democracy is that token holders are only digital representations of persons, and that a single entity or individual can use several digital representations[125] and therefore concentrate a multiple number of holdings, creating a potential dichotomy between crypto-asset holders' and users' interests. At the opposite, an off-chain governance model is advocated by Vlad Zamfir, the main technological architect of Ethereum[126].

Setbacks of on-chain governance have been described by Buterin: (i) a legitimacy issue as online voting by token holders would only reflect the views of a small percentage of people, and (ii) a non-cooperating attacker with only a small percentage of all tokens has the potential to sway the vote. Besides, in an on-chain governance model, DLT projects will have to be innovative to find incentives for investors to be involved and vote, as the individual capacity of token holders to influence the vote by their individual voice is small. All the more so as corruption of DLT online voting is possible: token exchanges could give interest rates to investors for deposits, and token holder profit maximizers could accept to vote in a certain way in return for payments from other token holders. But, on the other side, Buterin also warns that an off-chain consensus based only on core computer developers' votes is also the risk to have an "ivory tower" of specialists more interested by theoretical rather than economical and practical solutions for the DLT platform.

A third way has been proposed by Buterin in December 2017 in favor of "informal governance", as practiced by Bitcoin, Bitcoin Cash, Ethereum or Zcash. "Informal governance" would consist in a "multifactorial consensus", "where different coordination flags and different mechanisms and groups are polled, and the ultimate decision depends on the collective result of all of these mechanisms together. These coordination flags may include: the roadmap (i.e. the set of ideas […] about the direction of the project), the consensus among the dominant core development teams, the [token] holder votes, the user votes (through some kind of sybil-resistant polling system), the established norms."[127]

---

[122] Y-Y. Hsieh, "The Rise of Decentralized Autonomous Organizations: Coordination and Growth within Cryptocurrencies", Thesis, University of Western Ontario, June 2018; S. Davidson, P. de Filippi and J. Potts, "Disrupting governance: The new institutional economics of distributed ledger technology", SSRN, 19 July 2016, p. 17.
[123] I. Allison, "Kleros: Ethereum smart contracts meet ancient Greek legal democracy", IBTimes, March 2018: https://www.ibtimes.co.uk/kleros-ethereum-smart-contracts-meet-ancient-greek-legal-democracy-1665300
[124] F. Ehrsam, "Blockchain Governance: Programming Our Future", Medium.com, 27 November 2017.
[125] A single person can possess several crypto-asset wallets and therefore a plurality of digital identities.
[126] V. Zamfir, "Against on-chain governance", Medium.com, 1 December 2017.

Among the distributed governance models we could mention the Tezos model (an on-chain voting system where anyone can request a governance change, and if adopted a test period is necessary before the organization of a confirmation vote), the Dfinity model[128] (an on-chain voting system, and the possibility of direct and retroactive governance changes), the futarchy model[129] (DLT users would vote not on the implementation of particular policies, but about their general values, and prediction markets will use and implement the global policies that optimize such metrics, in order to address the "voter apathy" and behavioral irrationality issues), the holacracy model (a decentralized management governance based on self-governed and autonomous circles and roles for corporate stakeholders, "by disseminating decision-making power to autonomous and multi-disciplinary teams within an organization"[130]), the liquid or delegated democracy model (a part-direct and part-representative voting system where individuals vote directly themselves or delegate their right to vote to other individuals), the quadratic voting model[131] (voting system where voters may pay for votes, and the price of the votes acquired is the square of the number of votes acquired), the meritocratic governance models (i.e. the voting process takes into account the voters' reputation), while some academics are calling for "soft" decentralized governance[132].[133]

ii. Challenges and risks of the distributed governance models

The new models of distributed corporate governance are just in their infancy. Many new academic and business use cases will be experimented in the months and years to come, and we can only point to some challenges that these projects will have to focus about and address.

Reasons invoked frequently to adopt a distributed governance are to reduce the role of trusted third parties, to democratize corporate governance[134], to broaden transparency, to lessen information asymmetries and to minimize the risk of "tyranny of the majority"[135]. Yet consensus (or, more accurately, direct majority-based) decisions raise many questions from a governance perspective. Even though a crowd-sourced and majority-based decision-making is chosen to govern a given company, there is no existing empirical data that proves that investor involvement and free-rider problems would be significantly different than in a company governed by existing corporate governance rules based on the agent-principal and cost agency problematics. Distributed governance is based on "flat-hierarchy"[136] philosophy and on the idea that token holders will dedicate sufficient time to participate and vote according to the interests of the company. To strengthen investor involvement, many projects have thus developed incentives which can be summarized as a "monetization of attention" methodology, by granting rewards (tokens…) to acquire the attention of a network of investors[137]. What would be the powers allocated to these new kinds of company stakeholders such as smart contract developers, DLT users, oracles/smart contract validators, miners or curators? The role of these various stakeholders would need to be clearly identified and defined by DLT projects.

---

[127] V. Buterin, "Notes on Blockchain Governance", 17 December 2017; F. G'Sell, "The challenge of algorithmic governance", Interdisciplinary workshop on blockchains, Ecole Normale Supérieure, 2 July 2018.
[128] F. Ehrsam, "Blockchain Governance: Programming Our Future", Medium.com, 27 November 2017.
[129] V. Buterin, "An Introduction to Futarchy", Ethereum blog, 21 August 2014; R. Hanson, "Shall We Vote on Values, But Bet on Beliefs?", Journal of Political Philosophy, 2013.
[130] M. Fenwick and E. Vermeulen, "Technology and Corporate Governance: Blockchain, Crypto and Artificial Intelligence", ECGI Working Paper No. 424/2018, November 2018, p. 9.
[131] S. Lalley and E. Weyl, "Quadratic Voting: How Mechanism Design Can Radicalize Democracy", American Economic Association Papers and Proceedings, Vol. 1, No 1, 2018; S. Lalley and E. Weyl, "Nash Equilibria for Quadratic Voting", SSRN, January 2018; E. Posner, "Quadratic Voting", 30 December 2014; E. Posner and E. Weyl, "Voting Squared: Quadratic Voting in Democratic Politics", Vanderbilt Law Review, Vol. 68, No 2, 2015.
[132] B. Arruñada and L. Garicano, "Blockchain: The Birth of Decentralized Governance", SSRN, 11 May 2018.
[133] F. G'Sell, "The challenge of algorithmic governance", Interdisciplinary workshop on blockchains, Ecole Normale Supérieure, 2 July 2018; H. de Vauplane, "Blockchain et corporate gouvernance: stigmergie ou holacratie ?", Revue Banque n°821, 30 May 2018; F. Ehrsam, "Blockchain Governance: Programming Our Future", Medium, 27 November 2017.
[134] A. Wright and P. de Filippi, "Decentralized Blockchain Technology and the Rise of Lex Cryptographia", SSRN, 10 March 2015.
[135] C. Reyes, N. Packin and B. Edwards, "Distributed Governance", William & Mary Law Review Online, Volume 59, 2017; A. Tapscott, "Blockchain democracy: Government of the people, by the people, for the people", Forbes, 16 August 2016; D. Bradbury, "How Block Chain Technology Could Usher in Digital Democracy", CoinDesk, 16 June 2014.

Liability issues related to the use of DLT and smart contracts for managing companies are substantial[138]. Who would be responsible of a corporate decision made by a distributed governed organization, such as a DAO, without management, without a board of directors and decided by thousands of token holders? Would such unincorporated organization without a legal personality be recharacterized by a court or a national authority as a general partnership or a joint venture? What would be the applicable law or the competent jurisdiction? Who would be responsible for handling issues and disputes? Which liability rules would have to be provided to prevent/react to fraud, willful misconducts or malicious attacks (such as a computer crime[139], a or a 51% attack[140])? Who would pay in the event of an indemnification decision, and what kind of insurance would apply?[141] Besides, if a decentralized platform is not governed by a traditional incorporated and regulated company (e.g. decentralized token exchange platforms), how would regulators interact with such platforms? National laws and regulations will not be sufficient to solve these issues which are by definition transnational. International laws would have to be drafted on the matter, for example general principles about the contractual nature of transactions on decentralized platforms, about the specific instrument schemes for DLT users. Beyond, international jurisdictions may have to be created, for example a digital jurisdiction whose authority would be subject to an international treaty concluded between sovereign nations and whose governance would be difficult to establish.

In the meantime, corporate governance solutions may be found by DLT market participants themselves. To solve the frequent problem of the use of the funds raised by ICO funders, Ethereum founder Buterin has thought through innovative solutions, e.g. the "IICO"[142] and the "DAICO". Some ICO projects are based or this IICO model such as the Kleros project, or on this DAICO model such as "The Abyss" project, a digital distribution game platform based on a crypto-reward ecosystem[143]. The DAICO would limit the ability of ICO funders to use the funds raised in a way that does not satisfy the crypto-investors. This solution is not without risks: if the project funders hold a significant share of the tokens issued, they could potentially influence the token holders' vote to obtain the release of more funds. Besides, investors can also adopt a strategy of "free-riding" and "completely disengage by putting all their trust in the DAICO concept itself and therefore feel it is not necessary for them to actually participate in votes and resolutions, reducing the majority threshold and weakening the security of the mechanism."[144] That is the reason why a deep focus should be made on the token holders' financial and technological education.

---

[136] M. Fenwick and E. Vermeulen, "Technology and Corporate Governance: Blockchain, Crypto and Artificial Intelligence", ECGI Working Paper No. 424/2018, November 2018, p. 9.

[137] M. Field, "Decentralized Governance Matters", Medium, 5 February 2018.

[138] D. Zetzsche, R. Buckley and D. Arner, "The Distributed Liability of Distributed Ledgers: Legal Risks of Blockchain", EBI Working Paper Series 2017-007, No. 14, 15 August 2017; P. Paech, "The Governance of Blockchain Financial Networks", LSE Legal Studies Working Paper No. 16/2017, 16 December 2016.

[139] A. Barbet-Massin and J. Brosset, "La souscription de cryptoactifs et de jetons d'ICO : les recours des investisseurs", Revue Lamy Droit des Affaires, Supplement, RLDA 6531, September 2018, pp. 36-43.

[140] The "51% attack" refers to the event where the DLT provides a mining proof-of-work validation process of the transactions. If a single "miner" node possesses more than 51% of the DLT's mining/processing power, he may be able to manipulate the DLT code and insert in the chain of transactions fraudulent transactions, carry out a double transaction or even steal the crypto-assets. See: D. Guégan and C. Hénot, "A Probative Value for Authentication Use Case Blockchain", Documents de travail du Centre d'Economie de la Sorbonne, 2018.

[141] F. G'Sell, "The challenge of algorithmic governance", Interdisciplinary workshop on blockchains, Ecole Normale Supérieure, 2 July 2018.

[142] The "IICO" or "Interactive Initial Offering" is a token sale model which provides some advantages for minority investors (i.e. a personal investment cap to avoid excessive dilution, the possibility for investors to withdraw bids manually, no discount for institutional investors) and for entrepreneurs (i.e. an acceptance only of ether investment, a longer period of token crowd sale). See: C. Perreau, "L'interactive initial offering, l'alternative

There are also governance matters which have deeply divided the DLT community and may need some time before being subject to good practices and market recommendations, such as the use of forks. One main problematic with DLT is that rare events are often not properly assessed by decentralized communities (e.g. a hack in "The DAO" case). For classic financial transactions, it is possible to terminate or to amend them. In a DLT community, the main rules (about the confirmation of transactions, decentralization, consensus models…) are unwritten axiomatic rules agreed by the members/nodes of the community. As these intrinsic rules of the DLT are accepted implicitly[145] but not encrypted as such, in rare events (such as hacks, fraud or else), the entire set of rules (and the code) of a DLT could be modified in a couple of days or hours. In such rare scenarios, the question remains as to whether using forks is a good practice or a solution to avoid.

In computer science, a fork is "an open-source code modification"[146]. Soft forks are minor and retroactive modifications to the code without creating a new DLT, while hard forks consist in creating a new version of the DLT code different from the initial one. After the previously mentioned "The DAO" hack, two potential solutions were debated online by token holders: either to freeze the crypto-assets by updating any transaction initiated by the hacker's account (i.e. soft fork), which would have not necessarily required an approval by the vast majority of the token holders, or to modify the DLT code itself, in order to restore the original state of The DAO as it was before the stealing (i.e. hard fork), which would have required the agreement by a vast majority of the token holders votes, and which would have the risk to create profound disagreements. Hard fork was finally used in "The DAO" case in order to modify the transactions history by the community. This hard fork created a new DLT with a slightly different code than the initial one, which has allowed the investors to get back a large part of the stolen assets.

When an online community forks away from a crypto-asset it has decided to invest in, is it protecting the initial agreement between the members of this online community ("crowd-source venture capital") or is it a "social repudiation (or breach) of the "contract" itself"?[147] When the code of "The DAO" project has been hacked, the token holders have decided by a large majority to adopt the hard fork. It led to a big division, between those who wanted to intervene to eradicate the contested transactions, and those who wanted to respect the original code (no matter what its failures). It was the battle between the letter of the code versus the spirit of the

---

transparente à l'ICO", LeJournalDuNet.com, 26 October 2018; W. George, "Kleros' IICO Analysis", 26 July 2018; J. Teutsch, V. Buterin and C. Brown, "Interactive coin offerings", 11 December 2017.
[143] J. Halfon, "The DAICO: ICO savior or wolf in sheep's clothing?", Forbes, 24 May 2018.
[144] C. Pauw, "What is a DAICO, Explained", CoinTelegraph, 13 February 2018.
[145] Bitcoin.fr, "Hard fork/soft fork", 27 June 2016.
[146] Coindesk.com, "Hard Fork vs Soft Fork", 16 March 2018.
[147] A. Glidden, "Should Smart Contracts Be Legally Enforceable?", Blockchain at Berkeley, 27 February 2018.

code. It has been an important ideological and theoretical disagreement because it went to the heart of what constitutes DLT: do its users trust humans or the computer code? And is DLT a real trustless technology, without any interaction from humans and from third parties? In "The DAO" case, 87,28% of the token holders voted in favor of the hard fork[148], but what about the other 12,72% of investors which refused this proposal? And what would happen if, during a future similar hard fork vote, token holders which have voted in favor of the hard fork on the basis of certain assumptions that would appear ex post to be based on false or erroneous or fraudulent information (e.g. provided on Telegram or on another pseudonymous social media network)? Who would be responsible if one disagrees with the decision to fork?

Once the implementation of a fork is decided, the decision is final and cannot be technically retroverted. As provided by de Filippi, it is useful to provide that a social group is allowed to change or update at least its social contract, even though the latter is a smart contract. According to de Filippi, hard forks are the proof that "the ideal of a purely trustless technology is nothing more than an ideal".[149] The DAO vote proves that future human or technological interventions into the code of a DLT, and even a public one, could in the future be implemented ex post if such a decision is accepted on the basis of a majority decision by the token holders.[150]

## IV. Conclusion

The aim of this article is to provide an interdisciplinary analysis and contribute to the nascent legal and economic literature on the potential impacts of ICOs and tokenization on corporate governance. We have described the reasons behind the development of tokenization and fundraising methods using the distributed ledger technology.

Distributed ledger technology, its use cases (ICOs…), their regulation and their oversight are not mature and stabilized enough to conclude decisively on all the implications for companies that the DLT adoption by market participants could entail. Nevertheless, there is an increasing probability that new kinds of corporation stakeholders will emerge: the token holders. These new actors could lead to changes in the securities' issuance and trading, in the shareholder's involvement, but also to a reinforcement of the rights granted to the various corporation stakeholders. Beyond the focus only on shareholders, a new role could be attributed to corporate stakeholders, and to the online/client reputation of the firm.

We have also tried to emphasize what could be the emergence of a new perception of what constitutes a company, its management, its shareholders and its stakeholders. From a governance perspective, the path towards decentralization could not only impact the current state of corporate governance rules, and the nexus of contractual relationships within a firm, but also potentially on the actual definition of a firm. The role of the board of directors, for example, could be fundamentally disrupted and could potentially evolve to a mere supervision role of automated decisions made by computer technology, and even to its disappearance as there are no managers nor a board of directors in a DAO.

The possibility that a company would be governed by algorithmic code instead of human intervention (the so-called "algorithmic governance") is still at an early stage. Ethereum founder Buterin even referred this idea designed "to completely expunge soft mushy human intuitions and feelings in favor of completely algorithmic governance" as being "absolutely crazy"[151]. It is possible nevertheless that the evolution of corporate governance could be enhanced by algorithmic code.

---

[148] "Vote: TheDao Hard Fork", available on v1.carbonvote.com
[149] P. de Filippi, "La fin de l'idéal trustless", Blockchainfrance.net, 20 July 2016.
[150] D. Yermack, "Corporate Governance and Blockchains", Oxford Review of Finance, Volume 21, March 2017.

In the long term, based on a new kind of trust and on the belief that algorithms are more trustworthy than human intervention, it could be possible to establish governance rules at least partially in computer code, and to delegate the management of a company to algorithms which would automatically, under certain conditions, execute smart contracts and certain pre-determined actions.

But at this stage of knowledge of the DLT, the academic community lacks the knowledge and the theoretical tools needed to determine the real potential of distributed mechanisms to further our understanding of the tokenization process, of the DAO and of the distributed governance phenomenon. As previously indicated, the new models of distributed corporate governance are just in their infancy. Many new academic studies and business use cases will be experimented in the future, and we have tried in this paper to raise some of the legal and economic challenges that these projects will have to focus about and address.

---

[151] V. Buterin, "Notes on Blockchain Governance", 17 December 2017. See also: M. Atzori, "Blockchain Technology and Decentralized Governance: Is the State Still Necessary?", SSRN, 2 January 2016.